\documentclass[aps,twocolumn]{revtex4} 
\usepackage{amsmath,graphicx}
\newcommand{\andy}[1]{}
\newcommand{\tr}{{\textrm{Tr}}}
\newcommand{\mat}[1]{{\textbf{#1}}}

\begin{document}

\title{Multiple-photon resolving fiber-loop detector}

\author{J.~\v{R}eh\'{a}\v{c}ek} 
\author{Z.~Hradil}
\affiliation{Department of Optics, Palacky University, 17. listopadu 50,
772 00 Olomouc, Czech Republic}

\author{O.~Haderka}
\author{J.~Pe\v{r}ina~Jr.}
\author{M.~Hamar}
\affiliation{Joint Laboratory of Palacky University and the Institute of 
Physics of the Czech Academy of Sciences, 17. listopadu 50, 772~00 Olomouc,
Czech Republic}

\begin{abstract}
We show first reconstructions of the photon-number distribution obtained
with a multi-channel fiber-loop detector.
Apart from analyzing the statistics of light pulses this device
can serve as a sophisticated post-selection device for experiments 
in quantum optics and quantum  information. We quantify its efficiency 
by means of the Fisher information and compare it to the efficiency of the 
ideal photodetector.
\end{abstract}

\maketitle

A major drawback of common detectors of weak light fields
is their lack of photon-number resolution. Due to the nonlinear character
of amplifying process the response of such detectors is not 
sensitive to the strength of the input signal. The only
two detection events in such a case are ``click'' and ``no click'' 
that correspond to the presence or absence of the signal.
A device capable of photon-number resolution would contribute both
to fundamental research in quantum optics and to implementation of 
quantum communication and information protocols.
While such devices were recently indeed constructed 
\cite{kim99,kurtsiefer00,brattke01,santori01,pelton02,stevenson02},
they require operation under extreme conditions at present and therefore 
did not become a common laboratory tool yet. Also, their photon-number 
resolution is  still limited to only few photons. 

\begin{figure}
\centerline{\includegraphics[width=0.8\columnwidth]{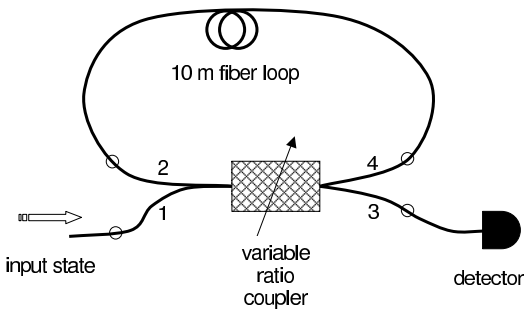}}
\caption{Fiber-loop detector \label{fig:loop}}
\end{figure}

Another way of circumventing this problem is splitting the input pulse
using a multiport device followed by an array of conventional binary 
detectors as was proposed in \cite{paul96,kok01}. In the ideal case 
of many output ports the
input pulse gets perfectly split and each photon is detected separately.
However reasonable performance of such device would require a very large number
of beamsplitters and detectors which would results in a bulky and costly
detection device. It has been suggested \cite{banaszek03}
to replace the complicated multiport device by a fiber loop and
a single photodetector, see Fig.~\ref{fig:loop}. 
After each round-trip part of the incoming pulse gets transmitted to a conventional 
binary detector. This results in a time resolved series of detections,
each of them corresponding to a different output port of the multiport device.
Such a  multiple photon resolving device has recently been built in our 
laboratory \cite{ondra03}. The variable ratio coupler inserted at the entrance
to the fiber-loop delay line, see 
Fig.~\ref{fig:loop}, is used to adjust the transmission
probabilities of the output channels to a certain extent 
and thus tweak the overall resolving power of our instrument.

The purpose of this communication is twofold. 
First, we will analyze the performance of the fiber-loop detector
and compare it to the efficiency of the ideal photodetection device. 
Second, we will show how to reconstruct the photon statistics of the input 
pulse from the data measured at the fiber-loop detector via the 
maximum-likelihood principle and apply this technique to experimental 
data.

The fiber-loop detector is used to count photons contained in the input
pulse in an indirect way. Therefore it is natural to relate its performance 
to that of the ``textbook photon counter''---the ideal photodetector
of quantum efficiency $\bar\eta$. The concept of the equivalent efficiency
is extremely useful in this context. The equivalent efficiency of the 
fiber-loop detector is defined to be equal to the quantum efficiency of the 
ideal photodetector that gives the same measurement error.   
This concept was used by Braunstein and Nemoto \cite{braunstein02} to analyze another
indirect photon-counting device---the homodyne detection scheme.
Since their motivation was to use the homodyne detection for signaling they 
chose the mutual information as their measure of the measurement accuracy.
Our motivation is different. We are rather interested in the 
analysis  of the input light statistics. Let $\rho_n$ be the true photon-number 
distribution (i.e. the diagonal elements of the density matrix) of the 
input pulse. After $N$ identically prepared pulses have been detected we can
make an estimate of the photon number distribution $\bar\rho_n$.
Due to the finite measurement time and the presence of noise it 
will usually differ from the true distribution. We will take the mean 
quadratic distance  $d=\sum_n(\rho_n-\bar\rho_n)^2$ between
$\rho_n$ and $\bar\rho_n$ as our measure of the estimation error. 
Let us note that for density matrices diagonal in the Fock basis  
this distance equals the Hilbert-Schmidt distance between
the true and estimated states.
Its reciprocal divided by $N$, $I=1/(d N)$, quantifies
the information about the 
input pulse gained from the measurement per pulse. Naturally, this distance 
depends not only  on the measurement device
but also on the chosen estimation strategy. Let us calculate
the error of the optimal estimation: It is well known that no matter how 
clever estimation procedure is adopted, its error cannot be smaller
than the Cram\'{e}r-Rao bound \cite{rao,cramer}. This bound applied to
$I$ implies that
\andy{inform}
\begin{equation} \label{inform}
I=\left[N\sum_n (\Delta\rho_n)^2\right]^{-1}\le 
(N\tr\mat{F}^{-1})^{-1},
\end{equation}
where $F$ is the Fisher information matrix,
\andy{fisher}
\begin{equation}\label{fisher}
\textrm{F}_{kl}=\left\langle\frac{\partial}{\partial\rho_k}\log 
P(\mathbf{n}|\rho)\frac{\partial}{\partial \rho_l}\log 
P(\mathbf{n}|\rho)\right\rangle_\mathbf{n},
\end{equation}
and $P(\mathbf{n}|\rho)$ is the conditional probability of
detecting the outcome $\textbf{n}$ provided the true statistics of the input
pulse is $\rho$. The averaging is done over the all possible outcomes $\textbf{n}$ 
of the measurement. The probability $P(\mathbf{n}|\rho)$ completely describes
the measurement apparatus. Provided the individual detection events
are independent, $P(\mathbf{n}|\rho)$ is a multinomial distribution,
\andy{multinomial}
\begin{equation}\label{multinomial}
P(\mathbf{n}|\rho)\propto\prod p_n^{N f_n}.
\end{equation}
Here $p_n$ is the probability that the input pulse
gives rise to a detection in a particular channel $n$.
The Fisher information matrix then simplifies to
\andy{fish-indep}
\begin{equation}\label{fish-indep}
\textrm{F}_{kl}=\frac{1}{N}\sum_n\frac{1}{p_n}
\frac{\partial p_n}{\partial\rho_k}
\frac{\partial p_n}{\partial \rho_l},
\end{equation}
\begin{figure}
\includegraphics[width=0.8\columnwidth]{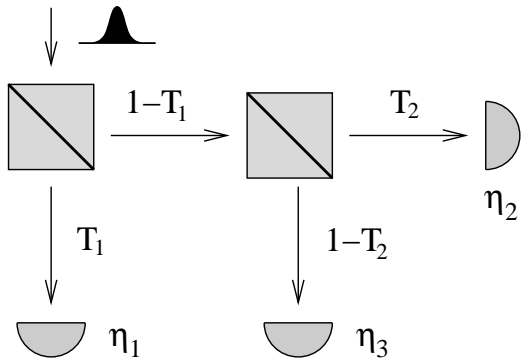}
\caption{Substituting scheme of the fiber-loop detector with three
output channels\label{fig:triple}}
\end{figure}
In the case of the ideal detector of efficiency $\tilde\eta$ the independent outcomes 
are just the different counted numbers of photons and the corresponding probabilities
are given by the well-known Bernouli distribution,
\andy{p-ideal}
\begin{equation}\label{p-ideal}
p_n=\sum_{m\ge n}\binom{m}{n}\tilde\eta^n(1-\tilde\eta)^{m-n}\rho_m.
\end{equation} 
Now we will analyze the fiber-loop detector which can be thought 
of as a multiport device. Such a multiport, for simplicity limited to only 
three output channels, is shown in Fig.~\ref{fig:triple}.

\begin{figure}
\centerline{
\includegraphics[width=0.75\columnwidth]{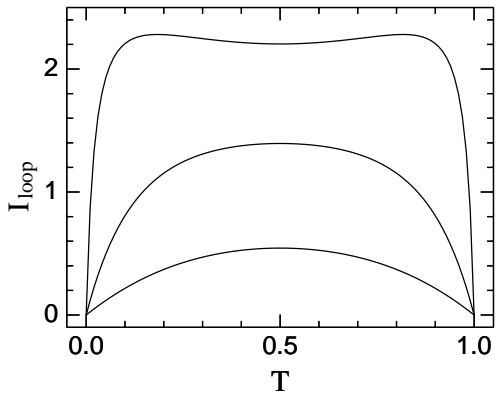}
}
\caption{Performance of the fiber-loop detector with two channels for
various input states and splitting ratios. Curves from below: 
$\rho_1=0.9$, $\rho_2=0.1$; $\rho_1=0.98$, $\rho_2=0.2$; $\rho_1=0.998$, 
$\rho_2=0.002$.
\label{fig:sedlo}}
\end{figure}

\begin{figure}
\centerline{
\includegraphics[width=0.75\columnwidth]{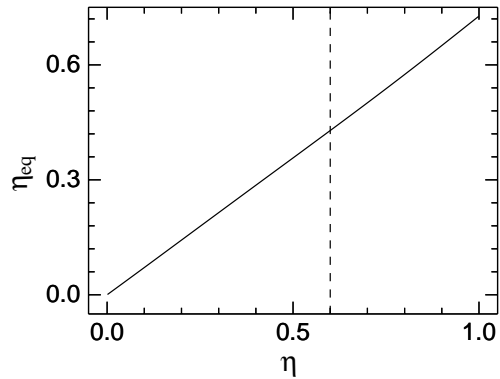}
}
\caption{Equivalent ideal detection efficiency of the fiber-loop detector
with three output channels in dependence on the quantum efficiency of the
binary detector used. The efficiency of standard commercial
single-photon detectors is about $60\%$ (dashed line). 
\label{fig:effective}}
\end{figure}

The simplest situation occurs when the coincidences between
detections at different output channels are not registered.
Then an $s$-channel device reduces to $s$ independent binary detectors
of generally different quantum efficiencies. As was shown in 
\cite{mogilevtsev98}, such a detection scheme already yields enough 
information for the reconstruction.
In this case the independent events are zero-detection events in the
individual channels having probabilities 
\andy{p-seq}
\begin{equation} \label{p-seq}
p'_j=\sum_m[1-\eta_j]^m\rho_m.
\end{equation}
A significant gain in accuracy of the reconstruction can be expected
when all possible coincidences are taken into account.
A detection event of such a complex observation is best recorded in the binary 
notation where ``1'' or ``0'' appearing at a certain position 
corresponds to a ``click'' or ``no click,''  respectively, in the given 
output channel. The most simple nontrivial case of such a measurement
is the fiber-loop detector with only two output channels (the pulse
goes round the loop only once). This is equivalent to letting 
the transmissivity $T_2$ of the second beamsplitter in Fig.~\ref{fig:triple}
go to unity. In this case we have four detection
events. Their probabilities read   
\andy{p-loop-2}
\begin{equation} \label{p-loop-2}
\begin{split}
p_{00}&=\sum_m\rho_m[1-\eta_1 T-\eta_2(1-T)]^m,\\
p_{10}&=\sum_m\rho_m[1-\eta_2(1-T)]^m-p_{00},\\
p_{01}&=\sum_m\rho_m(1-\eta_1 T)^m-p_{00},\\
p_{11}&=1-p_{00}-p_{10}-p_{01},
\end{split}
\end{equation}
and $T= T_1$.   
It is clear that only weak input signals can fully be analyzed by 
the simple measurement \eqref{p-loop-2}. In the following we will assume 
that  the input density matrix can be truncated and thus fully specified 
by giving its first few elements. The most simple case is
a pulse containing at most two photons,
\andy{0-1-2}
\begin{equation} \label{0-1-2}
\rho=(1-\rho_1-\rho_2)|0\rangle\langle 0|+\rho_1|1\rangle\langle 1|
+\rho_2|2\rangle\langle 2|.
\end{equation}
A successful estimation consist in identifying the single photon
and two photon contributions. Using Eqs.~\eqref{fish-indep}, 
\eqref{p-ideal}, and \eqref{p-loop-2} the information \eqref{inform} 
about the state \eqref{0-1-2} yielded by the fiber-loop
detector can easily be calculated. As the resulting expression
is rather complicated and not suitable for discussion we will consider
only the leading term of its expansions in $\eta$s. Even best
binary detectors have $\eta$ significantly smaller than unity
guaranteeing the rapid convergence of the series.   
Setting $\eta_1=\eta_2=\eta$ the information reads,
\andy{series-loop}
\begin{eqnarray}\label{series-loop}
I_\text{loop}&\approx&\frac{2\tau\eta^2}{5\rho_2}, \quad \rho_2\neq 0
\label{hill},\\
I_\text{loop}&\approx&\frac{2\eta}{\rho_1}+2(1-\tau)\eta^2,\quad \rho_2=0,
\label{shallow}
\end{eqnarray}
and $\tau=T(1-T)$.
As can be seen the information is determined by the  
last non-vanishing $\rho_n$. Naturally, states with large 
multi-photon contribution are more difficult to estimate.  
If $\rho_2$ significantly differs from zero it is optimal
to split the input pulse between the output channels equally, $T=1/2$,
see also Fig.~\eqref{fig:sedlo}.
However for states with negligible two-photon content $\rho_2\approx 0$ 
the information exhibits a local minimum at this point, see
Fig.~\eqref{fig:sedlo}. This minimum is caused by the term 
in Eq.~\eqref{shallow} which is quadratic in $\eta$ and since $\eta$ is
always significantly smaller than unity the minimum s a shallow one. 
In an experiment where the true 
state is not known, $T=1/2$ would be the best choice.  
The corresponding information gain of the ideal detector is 
\andy{series-st}
\begin{eqnarray} \label{series-st}
I_\text{ideal}&\approx&\frac{\bar\eta^2}{5\rho_2}, \quad \rho_2\neq 0
\label{ideal1},\\
I_\text{ideal}&\approx&\frac{2\bar\eta}{\rho_1},\quad \rho_2=0
\label{ideal2}.
\end{eqnarray}
As can be seen, for states with negligible two-photon content 
the ideal and fiber-loop detectors show the same performance.
This is not surprising because for single photon states 
there is no difference between the ideal and binary detector.
However for states with significant two-photon content the information
the fiber loop detector becomes  inferior to the ideal one.
The equivalent efficiency of the fiber-loop detector yields
in the worst case can be calculated by comparing Eqs.~\eqref{hill}
and \eqref{ideal1},
\andy{equiv-effic}
\begin{equation}\label{equiv-effic}
\eta_\textrm{eq}\approx\eta/\sqrt{2}.
\end{equation}
The dependence of the equivalent efficiency calculated by the exact 
inversion of the Fisher matrix on the quantum efficiency of the binary
detector used in the fiber-loop detector is shown in 
Fig.~\ref{fig:effective}. 
As one can see Eq.~\eqref{equiv-effic} makes 
an excellent approximation. Let us close this part noting that for weak input 
fields the information gain of the 
fiber-loop and ideal detectors differ only by the factor of $\sqrt{2}$.
Further improvement can be achieved by increasing the number of 
output channels. Provided the noise of the detectors can be neglected,
splitting output channels into two or more new channels
leads to a refinement of the probability operator measure describing
the detector. Any such refinement can only increase the information
gain. Such many-channel fiber-loop detectors will be discussed
elsewhere.

In real experiment experimenter's control of the parameters of the
fiber-loop detector is severely limited for many reasons, so it is not 
possible to choose the optimal configuration of the device.

Prior to its utilization the device has to be calibrated. We use coherent 
pulses whose statistics is known to be Poissonian for this purpose.
From the probabilities of zero counts at the individual output 
channels the transmissions of beamsplitters in the substituting scheme
are readily calculated. In the next step we calculate 
probabilities $p_n$ of all possible outcomes of the experiment.
We start from the probabilities of ``zero-arbitrary'' detection events.
Such events consist of detecting nothing in given output channels
while we do not care about the rest; they read
\begin{equation}
\tilde p_j=\sum_{k=0}^{\textrm{dim}}\bigl(1-\sum_{l=0}^{s}\epsilon_{jl} 
\eta_l\bigr)^k\rho_k,
\end{equation}
where $\text{dim}$ is the chosen cutoff of the Hilbert space
and $\epsilon_{jl}$ is zero if there was an arbitrary event 
(denoted $\forall$) detected at the $l$-th channel and unity if there 
was no detection, and $\eta_l=\langle n_l\rangle/N$ is the
detection efficiency of the $l$-th channel. 
Now the probability of having a coincidence at the 
first and third channels of a three-channel detector can be 
written in terms of the single-count probabilities as follows:  
$p_{101}=\tilde p_{\forall 0\forall}-p_{100}-p_{001}-p_{000}$. The latter
can, in turn, be expressed as $p_{100}=\tilde p_{\forall 00}-p_{000}$ 
and $p_{001}=\tilde p_{00\forall}-p_{000}$, respectively. 
The probabilities of all $2^s$ detection events possible with an 
$s$-channel loop detector can be easily calculated using this 
simple recursive procedure.  
Having $p_n$ determined one can proceed with the reconstruction.
Notice that $p_n$ are linear combinations of density matrix elements
$\rho_n$,
\begin{equation}\label{linpos}
p_j=\sum_i c_{ji}\rho_i, \quad j=1,\ldots,2^s,
\end{equation}
$c_{ji}=\partial p_j/\partial \rho_i$ being defined by the parameters
of the fiber-loop detector.  The problem \eqref{linpos}
is a typical linear and positive problem. The latter property follows
from the non-negativity of $\rho_n$.
Of course, the experimentally observed frequencies will generally differ 
from the theoretical probabilities, $f_j\neq p_j$.
Because of this the problem \eqref{linpos} has usually no exact 
solution and has to be solved in the statistical sense,
for example via minimization of some statistical measure of 
distance between $f_j$ and $p_j$. We use the approach based on minimizing 
the Kullback-Leibler divergence (also called relative entropy)
$d=\sum_j f_j \log p_j -\sum_j f_j \log f_j$ between 
data and theory. The state $\rho_n$ minimizing $d$ is at the same time the 
maximum-likelihood estimator of the input state and therefore attains the 
Cram\'er-Rao bound asymptotically for large $N$. 
The minimization can be carried on  e.g. via the iterative 
Expectation-Maximization algorithm \cite{dempster77,vardi93}:
\begin{equation}
\rho_k^{(n+1)}=\rho_k^{(n)} \sum_j 
\frac{f_j c_{jk}}{\sum_k c_{ji}\rho_i^{(n)}}.
\end{equation}
\begin{figure}
\centerline{
\includegraphics[width=\columnwidth]{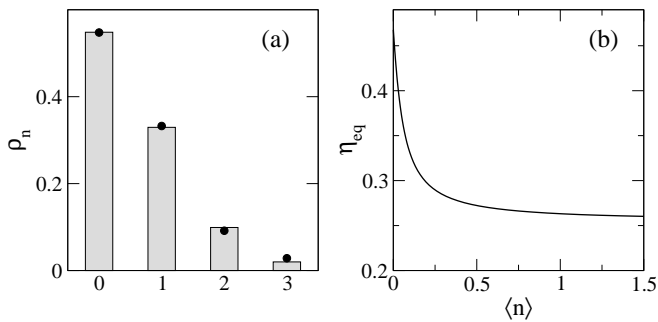}
}
\caption{(a) Reconstruction of the photon-number distribution. Black dots
are the best Poissonian fit with $\langle n\rangle\approx 0.6$ photons per
pulse. (b) The calculated equivalent efficiency of the fiber-loop detector
used in this experiment.
\label{fig:recon}}
\end{figure}
Left panel in Fig. \ref{fig:recon} shows the reconstruction of the photon statistics
of a laser pulse containing $0.6$ photons on the average.
The agreement between the measured and theoretical statistics is very good.
In this case only first three output channels were sufficient for the 
reconstruction. The equivalent efficiency of the three-channel
detector calculated for Poissonian light is shown on the right 
in Fig.~\ref{fig:recon}. 
We would like to emphasize that this is the efficiency of the real experimental
apparatus where a significant fraction of light is either lost inside
the apparatus or not detected due to small quantum efficiency of detectors
\cite{ondra03}.   
Nonetheless, in a wide range of intensities our detector is the equal of the ideal 
detector of quantum efficiency of about $25\%$.

\begin{figure}
\centerline{
\includegraphics[width=\columnwidth]{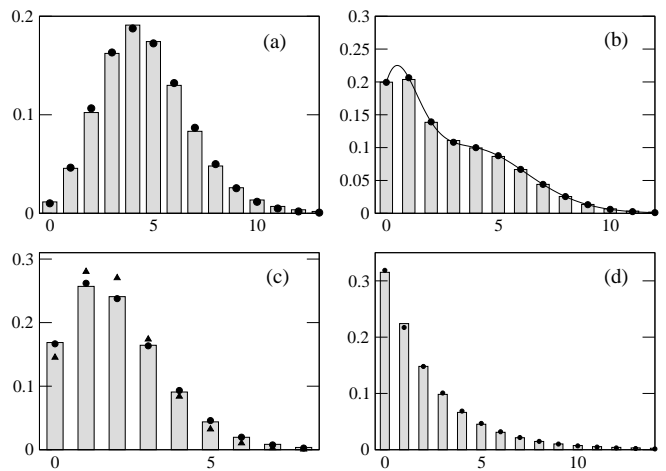}
}
\caption{Reconstructed photon-number distributions of different sources of 
light pulses
(bars); best theoretical fits (black dots); in panel (c) best Poissonian fit is shown
for comparison (triangles). Fluctuations:
(a) stable amplitude,  $P(I)\propto\delta(I-I_0)$;
(b) two-component amplitude, $P(I)\propto\delta(I-I_1)$+$\delta(I-I_2)$;
(c) uniform over the range of $[I_3-\delta,I_3+\delta]$;
(d) exponential, $P(I)=\exp(-I/I_4)/I_4$.
Resulting statistics: (a) Poissonian; (b) composite Poissonian;
(c) regularized generalized incomplete Euler gamma function;
(d) Bose-Einstein. Fitted parameters: $I_0=4.6$, $I_1=0.94$, $I_2=4.6$,
$I_3=2.0$, $I_4=2.1$, and $\delta=1.2$ photons per pulse.
\label{fig:multi}}
\end{figure}

More intense pulses
require the use of a larger number of output channels.
We have tested our fiber-loop detector on different kinds of light pulses.
The statistics has been artificially changed by changing 
the intensity of subsequent Poissonian pulses using a random number generator.
The statistics $P(I)$ of intensity fluctuations determines the 
true photon statistics of light pulses through the composition rule
\begin{equation}\label{compos}
\rho_n=\int \frac{I^n}{n!}e^{-I} P(I) dI.
\end{equation}
Experimental results obtained with different light sources 
are summarized in Fig.~\ref{fig:multi}.
Notice that the composite nature of $\rho$ nicely shows in the
upper right panel. Also the super-Poissonian character of the input
pulse is clearly seen in the lower left panel.
Several tens of thousands of pulses were used for each reconstruction.

In conclusion we have demonstrated the photon-counting capability
of the fiber-loop detector and quantified its efficiency 
by means of the Fisher information. 
The possibility to reconstruct the statistics of the input light pulses
of moderate intensities have also been shown.

This work was partially supported by grants LN00A015, and RN19972003012
of the Czech Ministry of Education, and by Czech-Italian project No.~29,
``Decoherence and Quantum Measurement.''

\end{document}